\documentclass{tdp}

\usepackage{graphicx}

\usepackage{amssymb}
\usepackage{amsthm}

\usepackage{amsmath}
\usepackage{algpseudocode} 
\usepackage{algorithm}
\usepackage{pdflscape}
\usepackage{array}
\usepackage[breaklinks=true]{hyperref}
\usepackage{breakcites}

\usepackage{booktabs}
\usepackage{tabularx}
\usepackage{ltablex} 
\usepackage{longtable}
\usepackage[small,bf,singlelinecheck=off]{caption}
\usepackage[section]{placeins}

\usepackage{geometry}
\usepackage{pdflscape}

\usepackage{subfig}
\usepackage{dblfloatfix}

\newcolumntype{a}{>{\arraybackslash}m{0.8cm}} 
\newcolumntype{f}{>{\arraybackslash}m{1.3cm}} 
\newcolumntype{k}{>{\arraybackslash}m{4.5cm}} 
\newcolumntype{i}{>{\arraybackslash}m{0.5cm}} 
\newcolumntype{L}{>{\arraybackslash}m{9.5cm}} 
\newcolumntype{A}{>{\arraybackslash}m{2cm}} 
\newcolumntype{T}{>{\arraybackslash}m{3.5cm}} 
\newcolumntype{S}{>{\arraybackslash}m{3cm}} 
\newcolumntype{V}{>{\arraybackslash}m{1cm}} 
\newcolumntype{P}{>{\arraybackslash}m{1cm}} 
\newcolumntype{Y}{>{\arraybackslash}m{1cm}} 
\newcolumntype{Z}{>{\arraybackslash}m{7.5cm}} 

\graphicspath{{Images/}}

\begin{document}

\title{Big Data Privacy Context: Literature Effects On Secure Informational Assets}
\author{Celina Rebello$^{a}$, Elaine Tavares$^{a,b}$}
\address{$^{a}$Federal University of Rio de Janeiro, Brazil - Coppead Graduate Business School \\ 
	$^{a}$Rua Pascoal Lemme, 355 - Ilha do Fundao - Cidade Universitaria, Rio de Janeiro - RJ\\ 21941-918 - Brazil.
	E-mail: $^{a}${\small \tt{rebello.celina@gmail.com}},$^{b}${\small \tt{elaine.tavares@coppead.ufjr.br}}
}
\TDPRunningAuthors{Celina Rebello, Elaine Tavares}
\TDPRunningTitle{Big Data Privacy Context: Literature Effects On Secure Informational Assets}
\TDPThisVolume{1}
\TDPThisYear{2018}
\TDPFirstPageNumber{1}
\TDPSubmissionDates{Received 24 August 2017; received in revised form 18 February 2018; accepted 23 April 2018}

\maketitle

\begin{abstract}
	This article's objective is the identification of research opportunities in the current big data privacy domain, evaluating literature effects on secure informational assets. Until now, no study has analyzed such relation. Its results can foster science, technologies and businesses. To achieve these objectives, a big data privacy Systematic Literature Review (SLR) is performed on the main scientific peer reviewed journals in Scopus database. Bibliometrics and text mining analysis complement the SLR. This study provides support to big data privacy researchers on: most and least researched themes, research novelty, most cited works and authors, themes evolution through time and many others. In addition, TOPSIS and VIKOR ranks were developed to evaluate literature effects versus informational assets indicators. Secure Internet Servers (SIS) was chosen as decision criteria. Results show that big data privacy literature is strongly focused on computational aspects. However, individuals, societies, organizations and governments face a technological change that has just started to be investigated, with growing concerns on law and regulation aspects. TOPSIS and VIKOR Ranks differed in several positions and the only consistent country between literature and SIS adoption is the United States. Countries in the lowest ranking positions represent future research opportunities. 
\end{abstract}


\section{Introduction} 
\label{sec:11}
\paragraph{}Big Data and privacy studies promote market excitement due to its perceived potential in research, business economy and social activities \cite{ref:gotterbarn_privacy_1999,ref:mcneely2014big,ref:Kshetri_2014}. Secure Internet Servers (SIS) are key data storage elements in big data's value chain \cite{ref:chen2012business,ref:van_de_pas_privacy_2015}. When individual's privacy enters the equation, frictional and also controversial effects show off like data misuse, user's overexposure, data breaches and many others \cite{ref:floridi_ontological_2005,ref:langheinrich2009survey,ref:tene2012big,ref:boyd_critical_2012}. One of these effects lays on gaps between current big data privacy theory and its practical indicators on key informational assets adoption like SIS. However, no study has evaluated big data privacy relation through Systematic Literature Review (SLR), bibliometrics, text mining approach and multi-criteria decision making methods (MCDM) rankings.

\paragraph{}This study main objective is the identification of research opportunities in the current big data privacy domain, giving a decision support alternative to researchers. First, the SLR provides the basis for bibliometrics mapping. Second, a theme and text mining analysis is performed over selected documents. Finally, the ranking on paper-country-SIS is performed with Technique for Order Preferences by Similarity to an Ideal Solution (TOPSIS) and the VlseKriterijumska Optimizacija I Kompromisno Resenje (VIKOR). 

\paragraph{}Big data privacy social sciences studies are focused on: users' concerns, awareness, self-management, self-disclosure, personality traits, privacy preservation. Variables as gender, age, education, and their relations have been explored but not exhausted \cite{ref:hull_contextual_2010,ref:bozdag_bias_2013}. The computational studies focus on encryption algorithms and informational security. Questions like ``how did they get my name?" \cite{ref:Culnan-1993} could be investigated with the support of research production versus informational assets indicators providing insights on big data privacy maturity level.

\paragraph{}Data is contextual, stored in servers, and desired by many \cite{ref:margulis1977conceptions,ref:nissenbaum2004privacy,ref:soria-comas_big_2015}. SIS are one of the data storage and transmission alternatives in the big data value chain. They are gateways to ``personally identifiable information", which makes them privacy violations targets just as safes in banks. They also reflects the investment level in Privacy Enhancing Technologies (PETs), used to protect stored data from unauthorized access. World Bank's SIS indicator from 2002 to 2015, in table \ref{tab:2}, provides a incipient, but valid, approximation to countries' big data privacy concerns. \cite{ref:posner1981economics}.

\paragraph{}Servers' violations have consequences like: impersonation, record misuse, patents theft and many other types of frauds. These effects impact economy and countries' sovereignty \cite{ref:Parsons1979171,ref:cukier2013rise}. The first privacy computational studies alert for encryptions and control at servers level \cite{ref:Va-1969,ref:MCCALMONTAM1970,ref:Young1970119} and on social impacts \cite{ref:DIALOE1970449} highlighting the relevance of stronger privacy regulations.

\paragraph{}The 21\textsuperscript{st} century computer was defined as ``an invisible technology aiming current world's improvement, transparently enhancing human (inter)actions" \cite{ref:weiser1999origins}. People are drown by sensors embedded in almost everywhere: cars, wearables and in-home appliances monitoring technologies. Physical barriers are over \cite{ref:konings2011territorial} and monitoring efforts are transparent \cite{ref:hartzog2013case}. Ubiquitous computing became big data's necessary environmental condition \cite{ref:diebold2012personal,ref:matzner_why_2014,ref:crampton_collect_2015}.

\paragraph{}In this study, section 2 presents the literature review that will provide SLR inputs for the search string. Section 3 describes the research method. Section 4 describes SLR results and TOPSIS and VIKOR ranks, eliciting research opportunities. Finally, section 5 discusses conclusions, limitations and implications.

\section{Literature Review}
\subsection{Privacy Definitions}
\label{sec:21}
\paragraph{}Privacy is conceived through many lenses. According to James F. Stephen, it is impossible to define privacy clearly. But, its violations are easy to point out. The intrusion of a stranger in someone's liberty be it by coercion, persuasion or even by the law are examples of privacy violations \cite{ref:stephen1991liberty}.
One of the first formal attempts to defy privacy is seen in Bentham's Inspection House. The Panopticon was designed as a place where an individual was set under custody and fully exposed \cite{ref:bentham2009panopticon}.
Its violations and dilemmas were first analyzed in 1890, when photography was the new technological phenomenon, and its use by journalists was ethically questioned. Privacy was defined as the ``right to be let alone"\cite{ref:warren_right_1890}.

\paragraph{}Other definitions are: ``claim of individuals, or groups, or institutions to determine for themselves when, how and to what extent information about them is communicated to others"\cite{ref:Westin1970}; ``the condition of not having undocumented personal knowledge about one possessed by others";``A person's privacy is diminished exactly to the degree that others possess this kind of knowledge about him. Documented information consists on information that is found in the public record or is publicly available" \cite{ref:parent1983privacy}; ``not simply the absence of information about us in the minds of others, rather is the control we have over information about ourselves" \cite{ref:Fried1984}, and ``the ability of the individual to control the terms under which personal information is acquired and used"\cite{ref:Culnan-1999}.

\paragraph{}Privacy is also defined in terms of protection from intrusion and information gathering, with individual control as choice, consent and correction \cite{ref:Tavani:2001:PPC:572277.572278}. Can also be stated as ``a right to control access to places, locations, and personal information along with use and control rights to these goods"\cite{ref:moore2008defining}.Other privacy perspectives are: combination of secrecy, anonymity, and solitude \cite{ref:gavison1980privacy}, and still  physical access, decisional, physiological and informational elements\cite{ref:tavani2008informational,ref:grodzinsky2010applying}. . Privacy is conceived as a value that, when present at some level, improves society relations in each and every term \cite{ref:Westin1970,ref:solove2002conceptualizing} \cite{ref:hull_successful_2015}; as a right that ought to be protected \cite{ref:warren_right_1890}; as a need to ensure liberty and autonomy\cite{ref:posner1977right,ref:posner1978privacy}. 

\paragraph{}Privacy can also relate to culture.This raises some questionings on its importance among all people, on what is inherently private or  merely social conventions \cite{ref:torra2017data}. Privacy definitions compilation revealed an overlapping among: (1) the right to be let alone; (2) limited access to the self; (3) secrecy; (4) control of personal information;(5) person-hood; and (6) intimacy \cite{ref:solove2002conceptualizing}. Privacy seems to be about everything, and therefore it is a vague concept. Still, a privacy taxonomy is based on informational processing, dissemination and violations\cite{ref:Solove2006}. In all cases, privacy faces limited protections by the law \cite{ref:warren_right_1890,ref:ambrose_right_2013}.

\subsection{Big Data}
\label{sec:22}
\paragraph{}Big data definitions have evolved in time and perspectives, from 3V's (Volume, Velocity and Variety), passing through 4V's (Veracity), 5V's(Variability), 6V's(Value) \cite{ref:gandomi_beyond_2015} definitions. Big data analytics operates on statistical methods and semantics extraction processes from both structured and unstructured captured data.
\paragraph{}When it comes to privacy, third parties data sharing and accessibility have a growing potential as investigation field \cite{ref:Kshetri_2014}. Theories such as Communication Privacy Management(CPM) \cite{ref:child2009blogging} and Privacy Calculus(PC) \cite{ref:dinev2006extended} are examples of data sharing studies which unveil new research opportunities focused on user self-exposure aspects.
 
\paragraph{}Both analytics and big data's capacities are associated in many definitions: automation, search, aggregation and cross geo-referentiation of massive data volumes \cite{ref:lyon2014surveillance,ref:boyd_critical_2012}. Big data applications intersect economical, strategical, security and consumer welfare domains \cite{ref:buhl_big_2013,ref:Kshetri_2014,ref:martin2015ethical,ref:zuboff_big_2015}, highlighting ethics as one of the most critical aspects \cite{ref:Zwitter2014,ref:Schroeder2014,ref:Qiu_2015,ref:hull_successful_2015,ref:selinger_facebooks_2015,ref:Taylor2016}.
\subsection{Big Data Practices}
\label{sec:23}
\paragraph{}Privacy threats affect not only law and computer science but social, psychology, economics and media studies \cite{ref:Culnan-1993,ref:heffetz_privacy_2014,ref:zuboff_big_2015}. Data breaches generate economical effects that cannot be ignored \cite{ref:spiekermann_challenges_2015}. 
\paragraph{}Literature have documented data records selling and information misuse practices \cite{ref:Culnan-1999}. These effects are negative among individuals, ranging from value destruction to rights violations, potentially jeopardizing big data's environment. People get surprised when they discovered about their data being used with unexpected purposes \cite{ref:Culnan-1993}. Complaints are mainly related to uninformed, not-consented and out of context data usage, generating data-context distortions and information abuse \cite{ref:nissenbaum2011contextual}. 

\subsection{Big Data And Privacy}
\label{sec:24}
\paragraph{}Big data privacy also seen as a surveillance dilemma \cite{ref:tene2012big, ref:lyon2014surveillance, ref:van2014datafication}. Big data constitutes a technological evolution with exponential scale effects, environmentally, constituted by ubiquitous computing \cite{ref:matzner_why_2014, ref:zuboff_big_2015} in a need to preserve individuals' privacy. If such balance is not reached, this environment will be jeopardized.

\section{Research Method}
\label{sec:31}
\paragraph{}Privacy literature reviews are non-systematic, many of them cannot be reproduced\cite{ref:DBLP:journals/access/Yu16,ref:DBLP:journals/misq/SmithDX11}. A few studies presented a systematic literature reviews related to security and privacy for big data, like \cite{ref:nelson2016security}. Hopefully, big data research count on Systematic Literature  Reviews (SLR)\cite{ref:frizzo2016empirical,ref:wamba2015big, ref:chen2012business}. Just a few studies are focused on classical bibliometrics indicators. However, there have been no big data privacy literature reviews providing research production analysis and  practical effects evaluation on computational assets. 
\paragraph{}This research is based on SLR method \cite{ref:kitchenham2004procedures}, added to a literature mapping exploring bibliometrics indicators. This work's objectives are: identify literature gaps ; analyze research themes and its evolution trough time; evaluate research opportunities per country.
The first two objectives are covered by the SLR with support of bibliometrics mapping and text mining. To evaluate research opportunities per country, TOPSIS and VIKOR were applied.

\paragraph{}This approach provides a precise, concise, technically reproducible and transparent evidence summary around a knowledge domain. The literature mapping elicits themes' evolution through time. Some of the mapped relations are: Most productive authors, countries, most related keywords, most cited authors, current research efforts and starting ones, themes' concentration areas and coupling relations.

\subsection{Journal Database, Search String And Data Collection}
\label{sec:32}
\paragraph{}Scopus was chosen as database because of its availability, broadness and reliability \cite{ref:Mongeon-2015}. A peer-reviewed research paper database, such as Scopus, provides a consistent platform to disseminate scientific investigation results, fostering research opportunities and trends.

\paragraph{}Chosen query parameters were ``Title-Abstract-Keywords"; limited to ``Articles" and ``Conference papers", written in English. ``Privacy" and ``Big data" queries intersect directly constraining the result from 2002 to 2016. Research string terms were chosen based on literature selection exposed in section 2. Exclusion criteria were: inaccessible, non-authored, and/or redundant documents.

\paragraph{}Search strings were constructed from Privacy and big data definition terms. First,privacy query revealed 83,657 publications (80,256 in English) while ``Big Data" returned  27,111 document results (26,076 in English). Search strings such as ``Priv*" included private funds and other themes that are out of the research scope. The intersection query retrieved 338 articles. The final string, which consists on the conjunction of privacy and big data search strings, returned 262 documents:

\paragraph{}\emph{TITLE-ABS-KEY ( ( privacy  AND  ( encrypt*  OR  crypt*  OR  auth*  OR  signature  OR  steganograph*  OR  anonymization )  AND  ( protect*  OR  secre*  OR  confident*  OR  "Polic*"  OR  control  OR  "self-management"  OR  preserv*  OR  hid* ) ) ) ) \textbf{AND} ( TITLE-ABS-KEY ( ( ( "Big Data"  OR  "Ubiquitous Computing"  OR  "Penetrate Computing" )  AND  ( "Informed Consent"  OR  disclos*  OR  expos*  OR  shar*  OR  distribut*  OR  dissemination  OR  "Data Exchange"  OR  "Data Trade" ) ) ) )  AND  ( LIMIT-TO ( DOCTYPE ,  "cp" )  OR  LIMIT-TO ( DOCTYPE ,  "ar" )  OR  LIMIT-TO ( DOCTYPE ,  "cr" ) )  AND  ( LIMIT-TO ( LANGUAGE ,  "English" ) )  AND  ( LIMIT-TO ( SRCTYPE ,  "p" )  OR  LIMIT-TO ( SRCTYPE ,  "j" ) )}. 

\paragraph{}In this sample, 13 documents had no authors and were excluded from the analysis. The final publication database has 226 publications. Articles with indexes 17, 199 and 249 had DOI errors. Thus, were not retrievable from Scopus.  Articles 51 and 52 are redundant and so are 261 and 262. Articles 30, 53, 106 and 107, 120, 139, 157, 198, 216, 233, 241, 255, 250, 254, 256 were not accessible either. Article 147 had only abstract available, which was removed from corpus. Article 253 was a talk in a book chapter, and only one page was retrievable.

\subsection{Information Retrieval And Classification Methods}
\label{sec:33}
\paragraph{}First, the documents went through bibliometrics mapping and analysis with the support of VOSViewer \cite{ref:van2009software}. Second, cluster and content analysis where employed to evaluate the key terms and their quantitative relevance. Hierarchical clusterization employed the \emph{k-means} method. On clusterization algorithms word stemming was included due to semantic aspects and theme's extraction objectives. So, words such as ``privacy" versus ``private" were also treated as a single keyword and resolved using wildcards, e.g. ``priva*".

\paragraph{}Text mining and classification were based on article contents inspection through \emph{Tf - IDf} matrix. All algorithms were operationalized in R language \cite{ref:R-Language} with support of``tm" \cite{ref:tm} and ``bibliometrix" \cite{ref:bibliometrix} packages. Terms relations, by similarities or distances, revealed concentration areas and relational gaps that were also confirmed by bibliometrics keyword analysis.

\subsection{Multi Criteria Decision Making Methods Ranking}
\label{sec:34}
\paragraph{}TOPSIS and VIKOR methods were chosen to check between practical literature effects on SIS. Both TOPSIS and VIKOR are widely considered as MCDM options\cite{ref:behzadian2012state}. Both methods could be applied to rank alternatives, propose a solution to the research question, having the decision criteria weights under decision-maker's discretion. Data convexity is not mandatory.
\paragraph{}TOPSIS purely employs analytical methods based on applying Euclidean distance functions on normalized vectors of positive (outputs) and negative (inputs) criteria\cite{ref:Hwang_and_Yoon_1981}. VIKOR determines the compromise ranking list, the trade-off solution, and the weight stability intervals of the obtained compromise solution \cite{ref:opricovic1998multicriteria}.

\paragraph{}TOPSIS and VIKOR focus on ranking alternatives selection in the presence of conflicting criteria. VIKOR provides a maximum ``group utility" for the ``majority" and a minimum of an individual regret for the ``opponent", its ranking index is based on the particular measure of ``closeness" to the ideal solution. TOPSIS rank has the ``shortest distance" to the ideal solution,  which is the  best level for all considered attributes, and the ``farthest distance" from the ``negative-ideal" solution, which is the one with worst attributed values. So, TOPSIS returns two ``reference" points, but it does not consider the relative importance of the distances from these points.
\paragraph{}TOPSIS and VIKOR proposal analysis are applied to generate rankings presented here as a publication impact alternative measure for big data privacy research and also point countries' consistency between papers' impact and SIS installed capacity. Ranks provide a supported decision mechanism to big data privacy researchers. 

\paragraph{}The percentage of non-cited-publications (NCP) was assumed as the negative ideal while all other criteria are considered positive. These ranks reveal a new approach on publication's relevance, differently than classical bibliometrics. Rankings provide big data privacy research indicator. 

\section{Results and Analysis} 
\label{sec:41}
\begin{figure*}[hbtp]
	\centering
	\captionsetup{justification=centering}
	\includegraphics[width=0.7\textwidth]{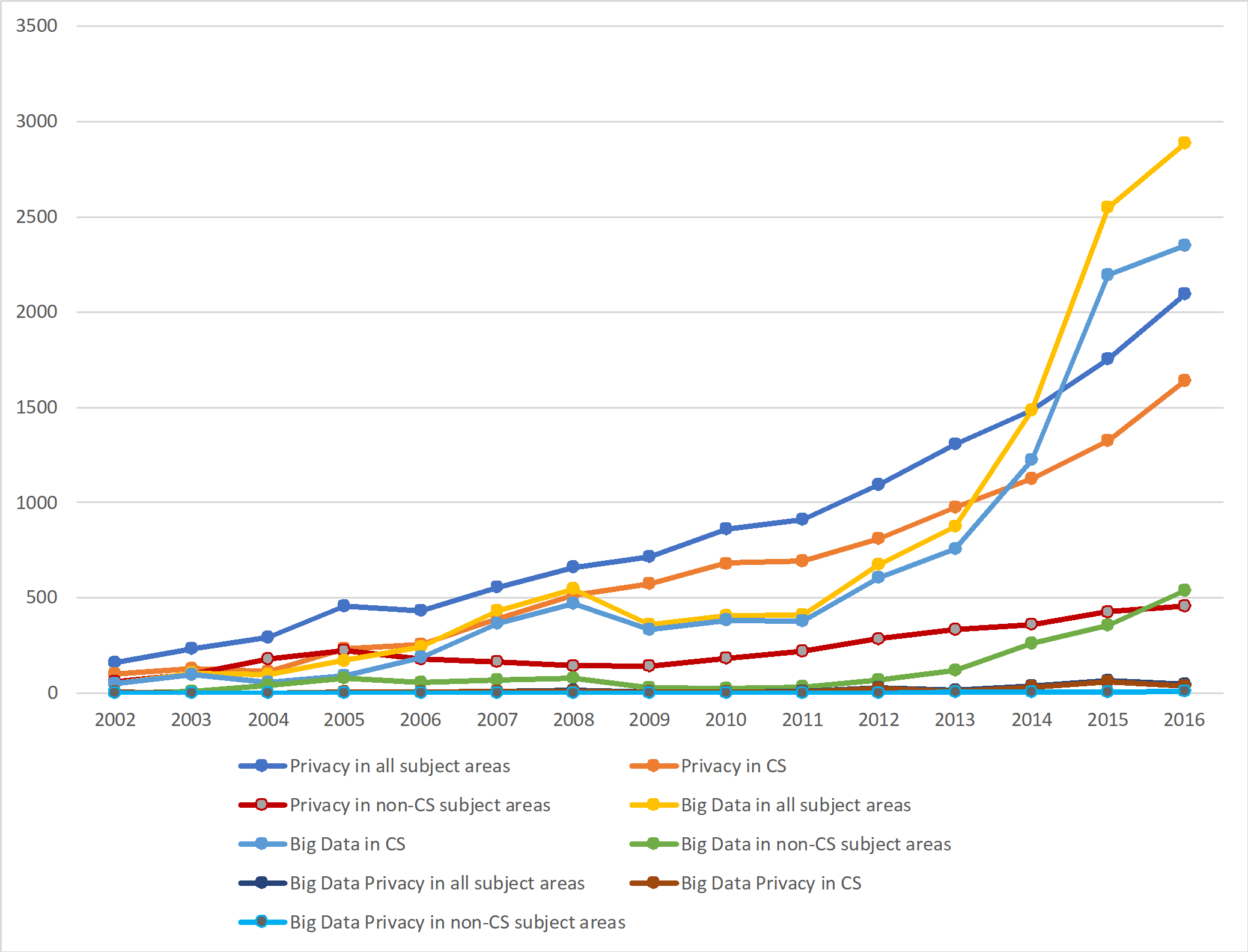}
	\caption{Big Data Privacy Themes And Research Evolution Per Year}
	\label{fig:1}  
\end{figure*}

\paragraph{}As presented on Figure \ref{fig:1}, the privacy and big data intersection had its first increase in 2008, with significant growth from 2013 to 2014, experiencing a Annual Percentage Growth Rate (APGR) of 18.614\% in all subject areas, having APGR of 16.644\% in computer science and 24.573\% in non-computer science areas.
\paragraph{}In the same period, privacy research had APGR of 21.762\% in all subject areas, 23.721\% in computer science alone and 17.035\% in non-computer science domains. Big data in all fields had 36.403\%, 34.670\% in computer science and 53.781\% in non-computer science areas. These rates reveals that other areas, different from computer science, turned their attention to big data. Big data privacy have taken other subject areas attention since 2005.

\begin{figure*}[hbtp]
		\centering
		\captionsetup{justification=centering}
		\includegraphics[width=0.58\textwidth]{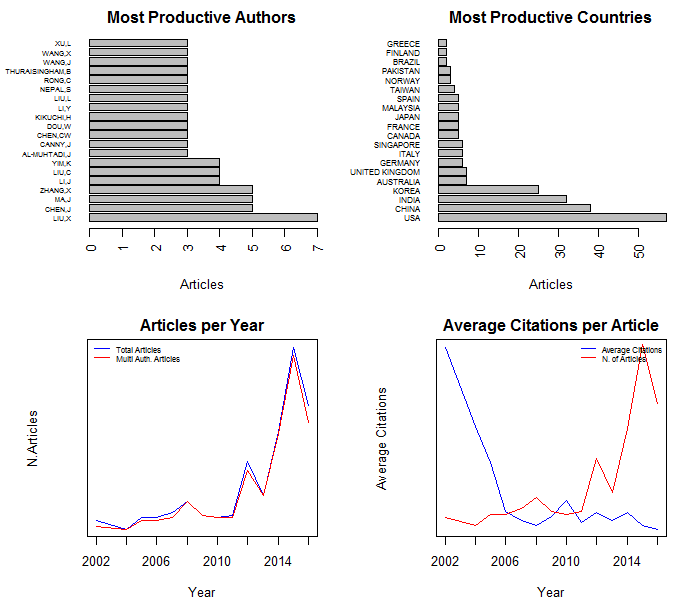}
		\includegraphics[width=0.58\textwidth]{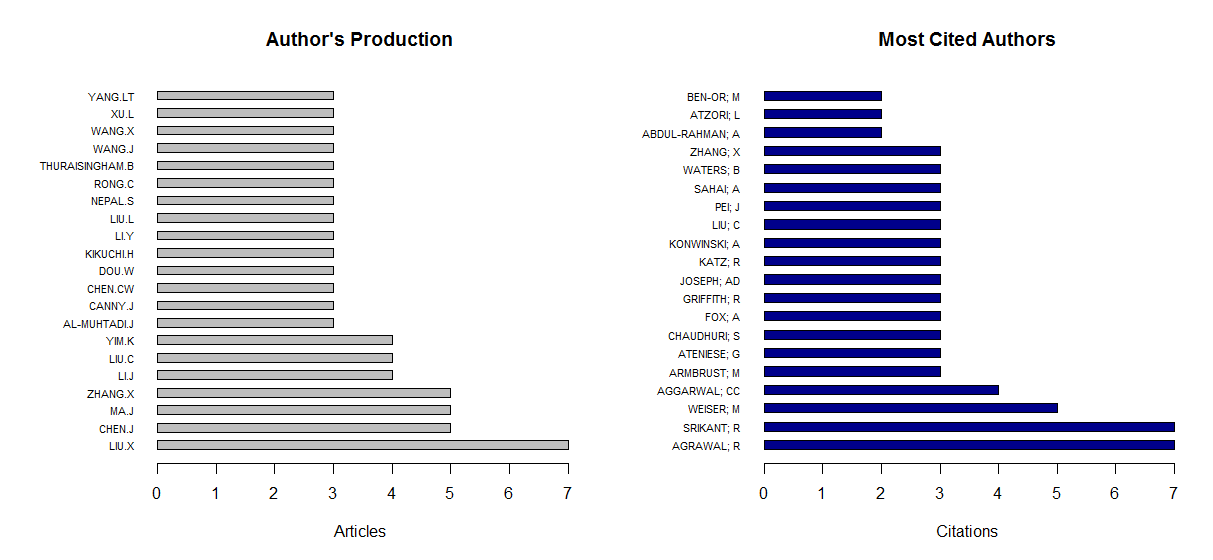}
		\caption{Bibliometrics Research Production Indicators}
		\label{fig:2}  
\end{figure*}

\paragraph{}Most productive authors are Chinese, associated to American institutions. Liu with 7 articles followed by Chen, Ma and Zhang with 5 articles. Most cited authors are Agrawal and Srinkant with 7 citations, and Weiser with 5 citations. Author's production shows Liu, and Zhang among most productive and cited authors, as seen in Figure \ref{fig:2}.

\begin{figure*}[hbtp]
		\centering
		\captionsetup{justification=centering}
		\includegraphics[width=0.58\textwidth]{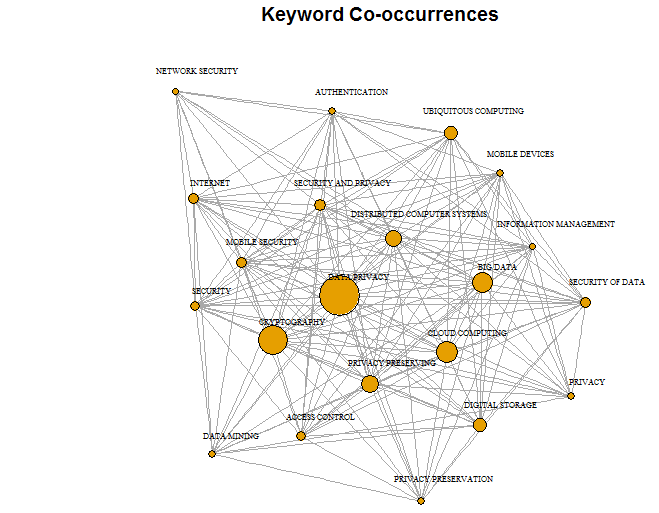}
		\caption{Top 20 Keywords Co-Ocurrences}
        \label{fig:3}    	
\end{figure*}

\paragraph{}Figure \ref{fig:3} shows Keyword Co-occurrence graph. Its edges reveal that``data mining" cannot reach ``access  control" directly. All minimum paths connecting these nodes have privacy or security related keywords, i.e. access control" can only be achieved if security as privacy aspects are considered. These relations indicate a new research challenge when non-digital aspects, such as ``shoulder surfing", and other off-line information gathering techniques are present.

\paragraph{} Most related keywords (Table \ref{tab:1}), returned Hadoop and MapReduce, not present in SLR search string. This indicates a potential relation between storage and file access technologies to privacy and to big data's Volume dimension.
\begin{table}[htbp]
	\caption{Most Related Keywords}
	\label{tab:1}
	\begin{tabular}{ikckc}
		\hline\noalign{\smallskip}
		& Author Keywords (DE) & Articles & Keywords-Plus (ID) & Articles \\ 
		\noalign{\smallskip}\hline\noalign{\smallskip}
		1 & PRIVACY                                      & 61 & DATA PRIVACY                                  & 92 \\ 
		2 & BIG DATA                                     & 41 & CRYPTOGRAPHY                                  & 61 \\ 
		3 & CLOUD COMPUTING                              & 27 & BIG DATA                                      & 46 \\ 
		4 & SECURITY                                     & 25 & CLOUD COMPUTING                               & 45 \\ 
		5 & AUTHENTICATION                               & 17 & PRIVACY PRESERVING                            & 41 \\ 
		6 & ACCESS CONTROL                               & 13 & DISTRIBUTED COMPUTER SYSTEMS                  & 32 \\ 
		7 & UBIQUITOUS COMPUTING                         & 13 & DIGITAL STORAGE                               & 29 \\ 
		8 & PERVASIVE COMPUTING                          & 12 & UBIQUITOUS COMPUTING                          & 23 \\ 
		9 & ANONYMIZATION                                & 10 & INTERNET                                      & 22 \\ 
		10 & DATA MINING                                  &  8 & MOBILE SECURITY                               & 22\\ 
		11 & PRIVACY-PRESERVING                           &  8 & ACCESS CONTROL                                & 21\\ 
		12 & ANONYMITY                                    &  7 & SECURITY OF DATA                              & 21\\ 
		13 & CLOUD                                        &  7 & SECURITY                                      & 20\\ 
		14 & ENCRYPTION                                   &  7 & DATA MINING                                   & 18\\ 
		15 & HADOOP                                       &  7 & DATA HANDLING                                 & 17\\ 
		16 & PRIVACY PRESERVATION                         &  7 & PRIVACY                                       & 17\\ 
		17 & CONFIDENTIALITY                              &  6 & SECURITY AND PRIVACY                          & 17\\ 
		18 & DATA ANONYMIZATION                           &  6 & SENSITIVE INFORMATIONS                        & 16\\ 
		19 & HOMOMORPHIC ENCRYPTION                       &  6 & PRIVACY PRESERVATION                          & 15\\ 
		20 & MAPREDUCE                                    &  6 & AUTHENTICATION                                & 14\\ 
		\noalign{\smallskip}\hline
	\end{tabular}
\end{table}

\begin{figure*}[htbp]
	\centering
	\includegraphics[width=0.65\textwidth]{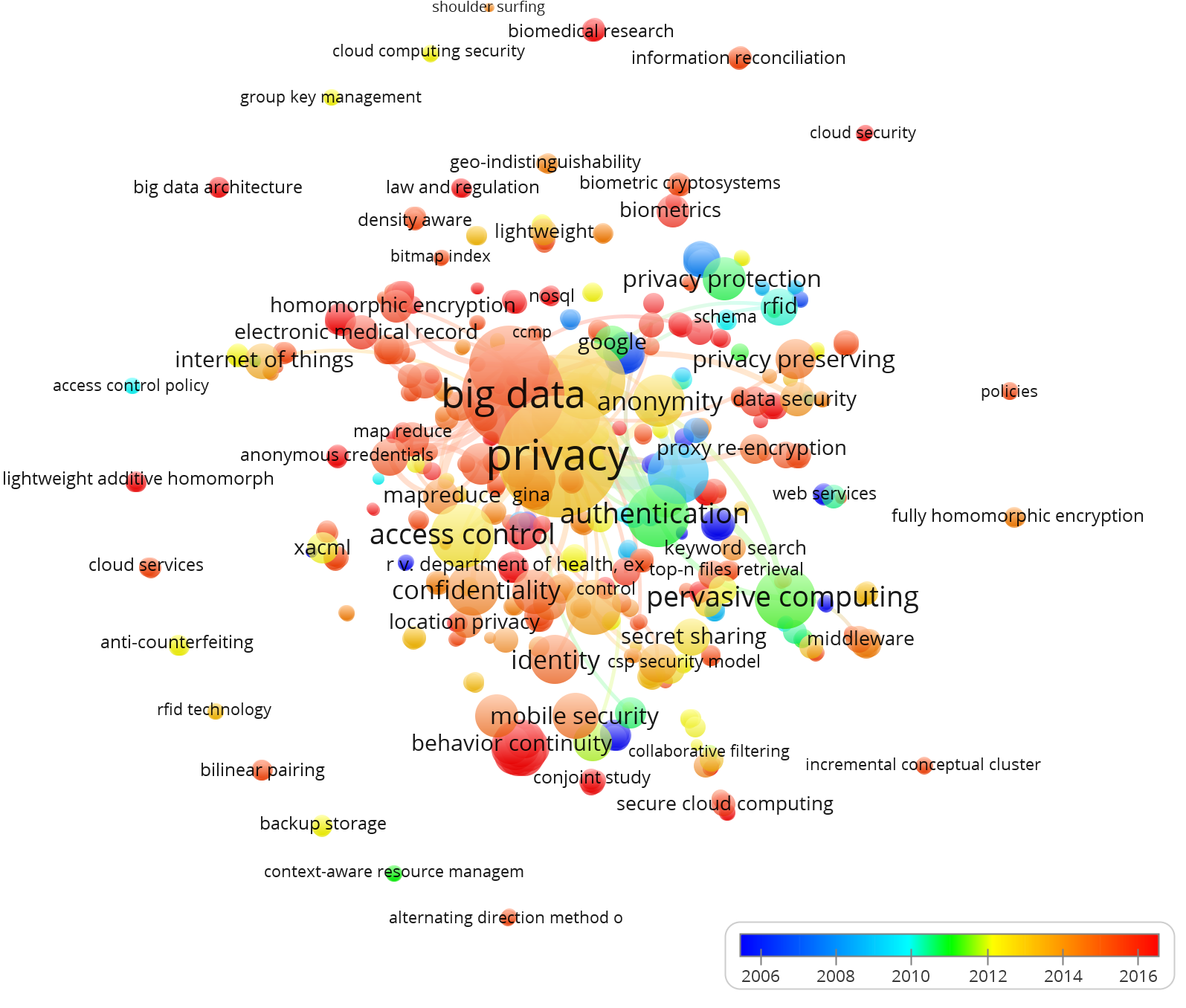}
	\caption{All Keywords Co-Ocurrences by Average Association Strength per Year}
	\label{fig:4}       
\end{figure*}

\paragraph{}Figure \ref{fig:4} shows that initial research was focused on ``access control policy", ``schema",``Proxy re-encryption". Access control policy defines which users or groups have permissions to access information. The proxy-re-encryption is a encryption process where a third-parties alter the previous encrypted cyphertext. These cryptosystems depends on ``schemes" and are relevant to protect user keys. All of these terms are related to SIS.
\paragraph{}Research production evolved to ``pervasive computing", ``authentication", ``privacy Protection", ``google", ``context aware resource management". Subjects like ``authentication" relates to ``servers" and also to ``access control", since the former depends on the later. As a research theme deployment, ``Privacy", ``anonymity", ``access control" ``homomorphic encryption", ``biometrics" became more relevant. 
\paragraph{}Keywords like ``secure cloud computing", ``incremental conceptual cluster" appeared as emerging research trends. Terms such as``law and regulation" are also under investigation and reasons are mainly because of ubiquitous computing effects. Other topics like "shoulder surfing" also called the attention because it is not related to big data itself, but as a information gathering off-line practice. 
\begin{figure*}[htbp]
	\centering
	\includegraphics[width=0.65\textwidth]{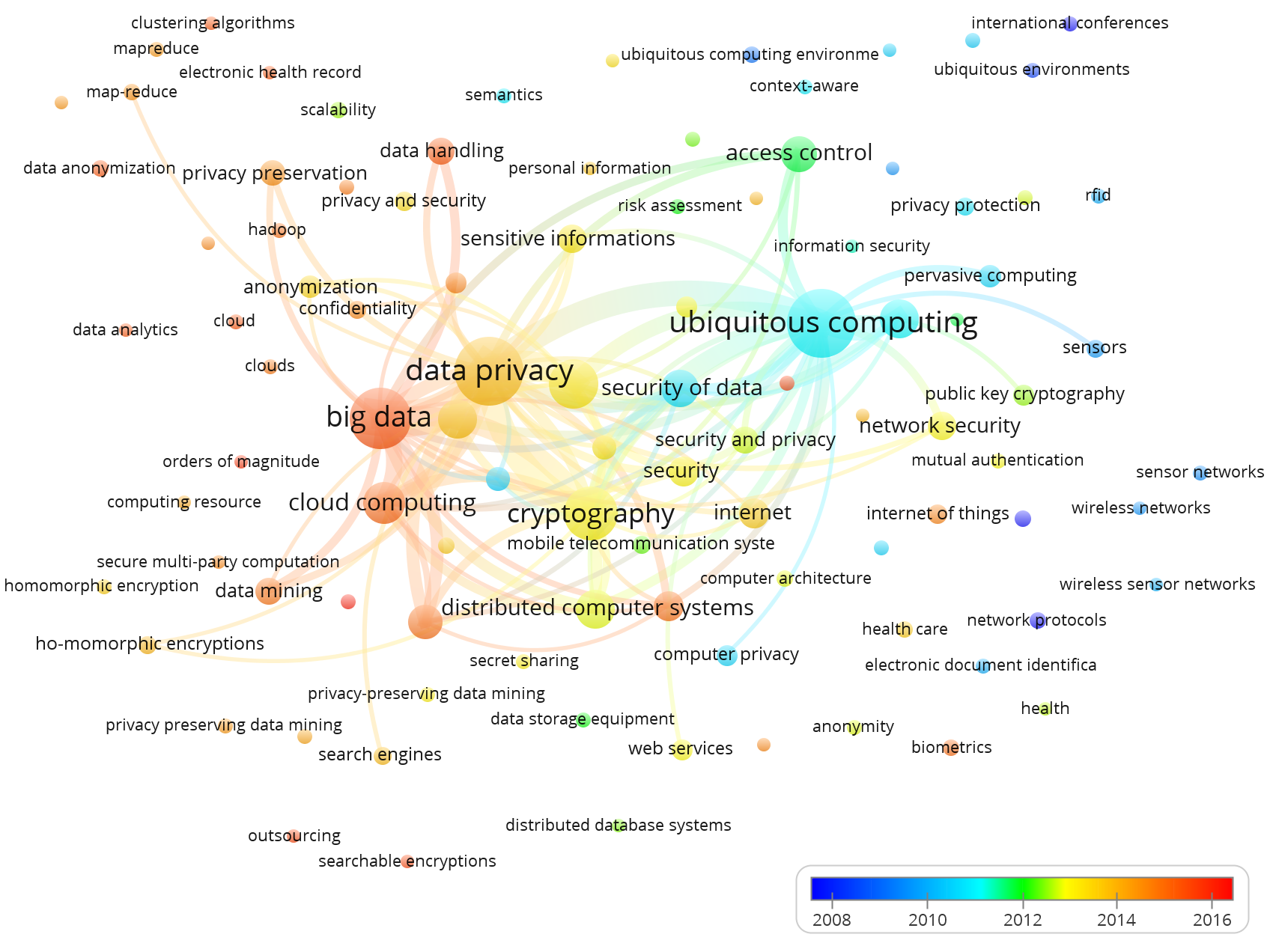}
	\caption{Top 100 Keywords Co-Ocurrences by Association Strength}
	\label{fig:5}       
\end{figure*}

\paragraph{}Keyword co-occurrence by association strength Figure \ref{fig:5} revealed research focus evolution from sensors, passing through ``access control" and finally reaching big data and privacy aspects. Such evolution is in conformity with what is perceived in \cite{ref:matzner_why_2014} when he states that privacy is never sufficient when computers are everywhere.
\paragraph{}Keyword co-occurrence by association strength also showed the themes' evolution from 2006 to 2016. The first stage research revealed concentrated efforts on: ``sensors", ``wireless networks", ``wireless sensor networks", ``context aware", ``semantics", ``computer privacy" and ``ubiquitous computing" as the most relevant among all of them.
\paragraph{}Research evolved from ``ubiquitous computing" to ``access control" and ``access control schemes" followed by ``scalability",``location" and ``data storage systems". The third research stage gathers ``data privacy", ``sensitive information", ``cryptography", ``anonymization", reaching ``big data", ``cloud computing", ``data handling" and ``data mining".
\paragraph{}Research themes started by sensors and networks, than evolved to scalability, storage and access issues. All these aspects are ubiquitous computing pillars. Later, works have focused on cryptography, privacy and security. Finally reaching ``big data", ``data mining", distributed and cloud computing caught researchers' attention. Privacy preservation aspects highlights the current research. 

\begin{figure}[htbp]
	\includegraphics[width=0.87\textwidth]{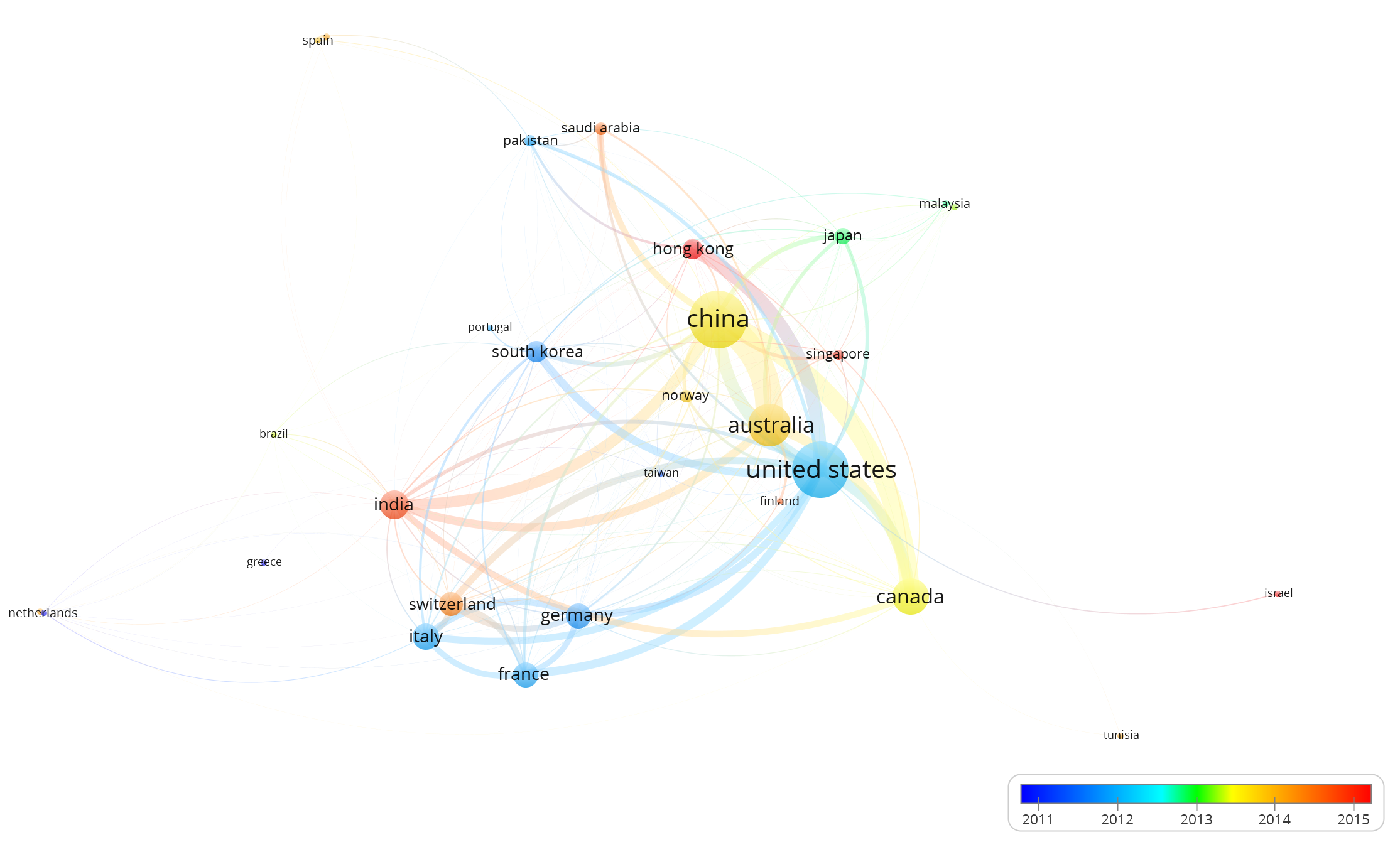}
	\caption{Bibliometrics Coupling per Country  association method - average normalized citations}
	\label{fig:7}       

	\centering
	\includegraphics[width=0.85\textwidth]{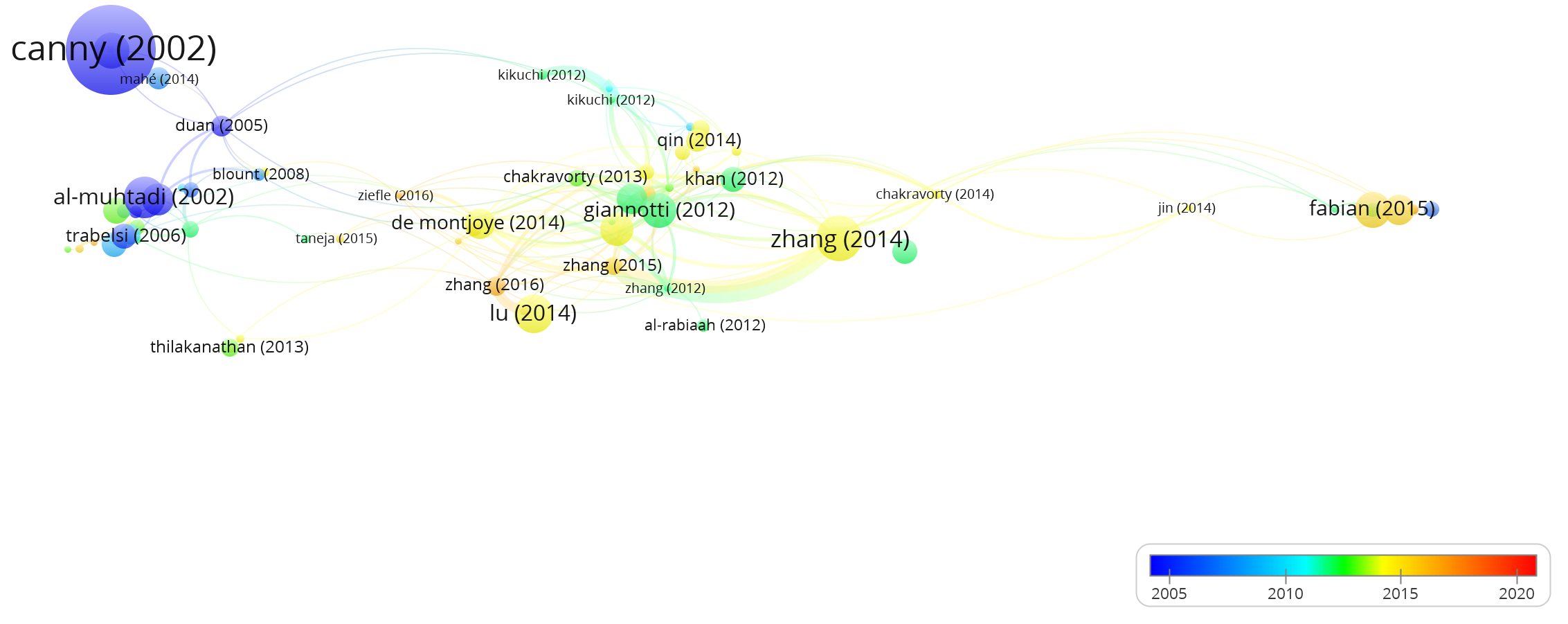}
	\caption{Bibliometrics coupling per documents}
	\label{fig:8}       

	
	\includegraphics[width=0.94\textwidth]{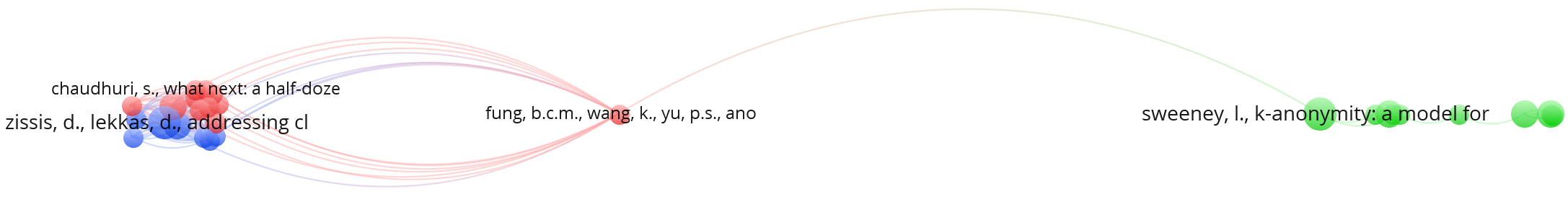}
	\caption[width=0.77\textwidth]{Co-Citations - with a minimum of 3 co-citations}
	\label{fig:9}       
\end{figure}

\paragraph{}Figure\ref{fig:7}, revealed 8 clusters, with minimum of one document per country and minimum one citation per document, with average normalized citations method. Canada, Germany and Saudi Arabia are leading countries in this metric. India, China and United States are leaders in research production. India leads in author's with most recent publications. 
\paragraph{}Figure \ref{fig:8} shows that bibliometrics coupling per documents, represented as (Author,Year), has a network of 100 references listed only once where 71 nodes and 10 clusters network. Canny's ``Collaborative Filtering with Privacy" and Al-Muhtadi's ``Routing through the mist: privacy preserving communication in ubiquitous computing environments" as most referenced articles in privacy intersection with big data research domains. These documents reinforce the SIS's role as a critical element in big data privacy research.
\paragraph{}Al-Muhtadi's work alerts to ubiquitous computing surveillance potential and proposes a \emph{``mist"} between routers. Canny's work defines a server-based collaborative filtering systems to protect people from monopolies. In this model, users control all of their log data. Users can compute a public ``aggregate”`
of all of their data without exposing individual users’ data. This model is based on homomorphic encryption with verification schemes distributed to all users. This is one of the first works to be proposed for untrusted servers. Both works propose privacy preservation through anonymization. Fabian's work on multi-cloud storage and sharing architecture is a natural evolution from both. This work focus on medical record anonymization shared among an cloud server array.

\paragraph{}Figure \ref{fig:9} shows three clusters, all related to storage, encryption and information security. In the first one, in red, literature mentions hadoop, mapreduce, privacy preserving and anonymization. The green cluster relates to privacy preserving, privacy enhancement,  anonymization, information classification. The blue one represents theme's convergence relation. Anonymization, privacy preservation are challenges in computer security. 

\clearpage
\subsection{TOPSIS and VIKOR literature effects evaluation}
\paragraph{}Table \ref{tab:2} shows several country research indicators. Countries research production, citations per paper and non-cited papers measure how relevant papers are and their impact. It presents research production frame against its practical effects and investments such as SIS per country. These differences reveal research opportunities on publication relevance, currently measured by citations. When these results are inserted on SIS adoption, there is an approximation between literature and its practical effects. 

\paragraph{}On citations per country the United States leads, followed by China, India and South Korea. United States also leads in average Citations per paper (CPP) indicator, followed by Canada and Germany. On SIS adoption rank results are different, with Switzerland leading with 3102 SIS, and China, India and Pakistan on the last positions.

\paragraph{}North American research is the most diversified and cited among all countries. It covers privacy awareness and preserving, surveillance and its economical effects privacy meta-data protection, network privacy architectures. Other topics are e-government policies, privacy usability challenges, self-disclosure, health, anonymization, geo-privacy, trust building and sensor networks.

\paragraph{}China has the second place in publications and in TOPSIS evaluation, and third in citations, with only 10 SIS. However, China comes in second on publications. Internet in China is strongly regulated \cite{ref:harwit2001shaping}, indicating that government controls its SIS. This condition reflects some potential difficulties on big data privacy research. Chinese publications relate to privacy preserving, trust building, authentication protocols, anonymity, encryption scalability and efficiency. 

\paragraph{}South Korea is forth in publications; second in citations, VIKOR and third in TOPSIS with 2320 SIS. Both Countries offer good research opportunities. South Korea's high SIS indicator would be explained by the companies' contribution in GDP, such as Samsung. Korean research production is one of the most diversified among all countries in the ranking. It varies from authentication and encryption schemes to information policies and e-government. User behavior aspects are rarely investigated.

\paragraph{}India is third in publications, forth in TOPSIS, last in VIKOR. India has the 19th position in SIS indicator. It is interesting to notice that many of Indian researchers are associated to American institutions, instead of Indian universities. This contributes to United Stated leading position. Big data privacy literature practical effects in India's are not as expressive as in other countries. Its research is concentrated on privacy preserving, anonymization algorithms, cloud computing, Internet of Things, wireless networks, health and trust building. User privacy awareness has not been investigated yet.

\paragraph{}Australian research is focused on privacy preserving, cloud, green and ubiquitous computing. Surveillance, trust building and big data sharing integrate the research production. This literature has close relation to the Canadian and Asian on privacy preservation.

\paragraph{}Canada's production is concentrated on privacy preserving through encryption algorithms and anonymization, access control and identity hiding schemes. Italian publications are on trust building in pervasive computing, anonymous mining, privacy preserving. Privacy law and regulations have not been investigated so far.

\paragraph{}United Kingdom's research focus varies from intrusion detection, Radio Frequency Identification secure based protocols, privacy systems for context aware and ubiquitous computing, wearables, transparency, ethics and health care. Research has interesting aspects because it focus from ethics to digital attacks countermeasures like intrusion detection systems. User's behavior and regulations were not explored in their production yet.

\paragraph{}French research is diversified and related to trust building, cryptography, biometrics, cloud and ubiquitous computing. Literature is marked by transparency on social mining and sharing. The trust building proposed by these works are in conformity with privacy by design fundamentals. Trust related papers are the most cited among the French research production.

\paragraph{}German studies include wearables, willingness to share, big data privacy health care management, social mining, trust in ubiquitous computing. These studies are mostly focused on building and ensuring trust. Their fundamentals are aligned to privacy by design principles \cite{ref:d2015privacy}.

\paragraph{}Japanese papers focus on privacy preserving aspects and challenges. These studies may integrate a new framework that addresses changing business needs and fresh concerns over breaches of personal data. Asian are marked by a strong privacy preservation bias. Regulation development and privacy breaches effects are research opportunities in the country as well as in the Asian region.

\paragraph{}Singapore's research production is on privacy preserving. Singapore enacted the Personal Data Protection Act (PDPA) in 2012, and in 2014 it became fully operational. PDPA's effects on corporate governance and data protection practices are also research opportunities in the country and region.

\paragraph{}Malaysian publications are diversified from soft computing, privacy concerns on health records and intrusion detection. Malaysia invests in high end engineering research development. Privacy and data protection by design principles are also an opportunity with potential positive effects on country's strategy \cite{ref:danezis2015privacy}.

\paragraph{}Saudi Arabia presents works focused essentially on cloud computing frameworks and ubiquitous computing. Papers application domains are health and multimedia, with utility driven to policy making. These studies take into consideration current frameworks security flaws on their analysis. Ubiquitous computing plays a key role on Saudi Arabia diversification strategy: from oil to big data. Thus, further big data privacy research is a need.

\paragraph{}Spanish papers vary from anonymization to PETs, passing through surveillance and authorization. Interesting to notice that user privacy awareness was absent as theme. This is also an aspect to be investigated as research opportunity.

\paragraph{}Norwegian works are related to anonymization and privacy preserving. These works present as fundamentals some of the principles present in \cite{ref:d2015privacy} like transparency and control. Further developments can be derived from these works, contributing to privacy by design philosophy through the proposed frameworks.

\paragraph{}Pakistan's research production has a strong focus on protocols, frameworks and signature schemes to build trust. This country has no privacy laws nor policies. This affects country's social and economical development. This is also a research opportunity.

\paragraph{}Switzerland appears 18th in publications, 4th in citations per paper, 8th in TOPSIS and 9th in VIKOR, and leads with 3102 SIS. It is interesting to notice that: Switzerland is considered a financial center, it has a high SIS indicator, and also a high Citation per Paper indicator. Swiss citations per paper is the forth highest in the rankings. A new research opportunity is on measuring if and how financial organizations influence affects this relation.

\paragraph{}Brazilian research production focus on cloud computing security and privacy framework development. These efforts complement the PETs works and also protocols. Brazilian research did not focus on big data social aspects yet. Big data privacy effects on human behavior may become a prolific opportunity, like anti-fraud detection and also government transparency to fight corruption.

\paragraph{}Netherlands is in 20th place regarding number of publications, but 5th in TOPSIS and 7th in VIKOR. Difference between TOPSIS and VIKOR proposed rankings can be due to normalization method applied in these methods. Netherlands' high relation between Citations per Papers and SIS indicator is a research opportunity on how this relation is established. Netherlands' research production is mainly focused on information security. There are still several applications to be covered, specially in ubiquitous computing and trust building. 

\paragraph{} There is a discrepancy between number of publications and SIS on countries like Brazil, Spain, Malaysia, Saudi Arabia and United Kingdom. These countries have a small research production, less than 10 publications in the analyzed period,  and represent new big data privacy local relations to explore. The small publication number indicates research venues, unexplored local opportunities. Big data privacy questions, specially in law and regulation are still concealed. Furthermore, both analysis help in identifying the publication efficiency and effects on SIS implementations and data breaches. 

\pagebreak
	\begin{table}[!htbp]
	\resizebox{0.63\textwidth}{!}{\begin{minipage}{\textwidth} 
		\centering
		\caption{Bibliometric Indicators Ranking Compared To TOPSIS And VIKOR} 
		\label{tab:2}	 
			\begin{tabular}{alcccccccccccccc}
				\hline\noalign{\smallskip}
				& Country & Pub & Cites & CPP & Std.Dev & NCP & Max.Cites & Pub.SIS & SIS & T.s & T.r & V.S & V.R & V.Q & V.r\\ 
				\hline\noalign{\smallskip}
				1 & United States & 67 &  857 & 12.791 & 29.834 & 0.388 & 207 & 0.041 & 1650 & 0.667 & 1 & 0.157 & 0.090 & 0.000 & 1 \\ 
				2 & China & 46 &  197 & 4.283 & 10.114 & 0.500 & 54 & 4.600 &   10 & 0.428 & 2 & 0.722 & 0.142 & 0.872 & 12 \\ 
				3 & India & 33 &   33 & 0.971 & 2.634 & 0.618 & 15 & 4.714 &    7 & 0.341 & 4 & 0.909 & 0.143 & 1.000 & 20 \\ 
				4 & South Korea & 29 &  200 & 6.897 & 24.823 & 0.448 & 134 & 0.012 & 2320 & 0.419 & 3 & 0.482 & 0.111 & 0.412 & 2 \\ 
				5 & Australia & 16 &  102 & 6.375 & 13.266 & 0.375 & 54 & 0.011 & 1460 & 0.249 & 9 & 0.668 & 0.127 & 0.691 & 5 \\ 
				6 & Canada & 10 &  126 & 12.600 & 18.422 & 0.400 & 54 & 0.008 & 1309 & 0.298 & 7 & 0.590 & 0.127 & 0.641 & 3 \\ 
				7 & Italy & 10 &   81 & 8.100 & 11.120 & 0.300 & 31 & 0.035 &  289 & 0.203 & 13 & 0.727 & 0.131 & 0.764 & 6 \\ 
				8 & United Kingdom & 9 &   31 & 3.444 & 8.487 & 0.444 & 26 & 0.007 & 1383 & 0.165 & 16 & 0.794 & 0.139 & 0.887 & 13 \\ 
				9 & France & 8 &   57 & 7.125 & 11.180 & 0.250 & 31 & 0.010 &  813 & 0.207 & 12 & 0.712 & 0.135 & 0.791 & 8 \\ 
				10 & Germany & 8 &   93 & 11.625 & 14.481 & 0.125 & 36 & 0.005 & 1757 & 0.298 & 6 & 0.559 & 0.132 & 0.662 & 4 \\ 
				11 & Japan & 8 &   11 & 1.375 & 0.518 & 0.000 &  2 & 0.008 &  971 & 0.201 & 14 & 0.796 & 0.143 & 0.925 & 17 \\ 
				12 & Singapore & 6 &   41 & 6.833 & 14.825 & 0.500 & 37 & 0.006 &  932 & 0.196 & 15 & 0.753 & 0.137 & 0.844 & 10 \\ 
				13 & Malaysia & 5 &   29 & 5.800 & 11.323 & 0.400 & 26 & 0.048 &  104 & 0.155 & 18 & 0.809 & 0.139 & 0.900 & 14 \\ 
				14 & Saudi Arabia & 5 &   27 & 5.400 & 4.219 & 0.200 & 11 & 0.093 &   54 & 0.157 & 17 & 0.816 & 0.140 & 0.915 & 15 \\ 
				15 & Spain & 5 &    8 & 1.600 & 2.510 & 0.400 &  6 & 0.014 &  362 & 0.084 & 20 & 0.909 & 0.143 & 1.000 & 19 \\ 
				16 & Norway & 4 &   14 & 3.500 & 1.291 & 0.000 &  5 & 0.002 & 2033 & 0.238 & 11 & 0.724 & 0.142 & 0.867 & 11 \\ 
				17 & Pakistan & 4 &   22 & 5.500 & 1.915 & 0.000 &  8 & 2.000 &    2 & 0.244 & 10 & 0.787 & 0.143 & 0.919 & 16 \\ 
				18 & Switzerland & 4 &   35 & 8.750 & 14.953 & 0.500 & 31 & 0.001 & 3102 & 0.289 & 8 & 0.639 & 0.141 & 0.799 & 9 \\ 
				19 & Brazil & 3 &   13 & 4.333 & 6.658 & 0.333 & 12 & 0.039 &   77 & 0.124 & 19 & 0.852 & 0.143 & 0.962 & 18 \\ 
				20 & Netherlands & 3 &   25 & 8.333 & 10.214 & 0.000 & 20 & 0.001 & 2828 & 0.301 & 5 & 0.575 & 0.143 & 0.778 & 7 \\ 
				\hline
			\end{tabular}
	\end{minipage}}
	\end{table}
\resizebox{0.63\textwidth}{!}{\begin{minipage}{\textwidth}
    Pub: Number of Publications; Cites: Number of Citations\\
	CPP: Citations per paper; Std.Dev: Citations Standard Deviation\\
	Max.Cites: Maximum Citations\\
	NCP: percentage of Non-Cited Papers\\
	T.s:TOPSIS Score; T.r:TOPSIS Ranking\\
	V.s:VIKOR Score; V.r:VIKOR Ranking\\
	Pub.Sis: Publications/SIS\\
	SIS: Secure Internet Servers in 2015 \\ Source:http://data.worldbank.org/indicator/IT.NET.SECR.P6?view=chart\\retrieved in 13/12/2016.
	\end{minipage}}

\section{Conclusion}

\paragraph{}This study consists on a RSL, bibliometrics mapping and text mining analysis on big data privacy research evaluation. TOPSIS and VIKOR MCDM Methods were employed to evaluate research practical effects, identifying new research rankings and opportunities.

\paragraph{}Privacy in big data is richly represented in the computer science domain, but non-computer science areas have started to investigate it. This study identified ``access control", anonymization, authentication and PETs as recent concentration areas and also ``ubiquitous computing" as a necessary environmental condition to big data. Non-computer science studies are concentrated on privacy perservation, trust building and privacy self-management. Computer Science studies are focused on encryption, anonymization, storage, cloud computing and data mining. 

\paragraph{}The TOPSIS and VIKOR Rankings revealed that United States leads on research impact and on the applying literature practical effects, which are represented by SIS. SIS ranking per country was the chosen criteria because it is an worldwide accepted computational asset indicator available from the World Bank. Another reason was SIS technical essence, which is secure data storage and transmission. 

\paragraph{} Rankings revealed that countries like Brazil, Spain may represent new opportunities according to both rankings. Saudi Arabia and United Kingdom, India, Japan and Pakistan according to VIKOR ranks. It is interesting to notice that Asia and Europe have research bias, driven to ethical aspects and privacy preserving, while United States drives efforts towards encryption, storage, and technical frameworks. Arabian countries investigate themes related to their economical growth. Latin countries like Brazil have just started to research big data privacy. Countries with a incipient research production is prolific in investigation opportunities because too little is known out their reality and matters. 

\paragraph{}Results may vary if inclusion and exclusion criteria are changed. Ranking may also change according to chosen MCDM method, criteria and weights adopted by decision makers. Since there was no previous study relating research production and MCDM methods, this work adds a contribution on a structured process where researchers should focus their efforts. SIS can provide non-exhaustive, but still relevant, measure of privacy concern per country. It is massively present in computer science research production and represents a key factor in data protection and application services.

\paragraph{}The whole process had to be documented, including intermediary results to avoid inconsistency. Data retrieval depended on Scopus' search engine technical structure. Article's classification by publishers is a biased process and another recognized limitation. Documents exclusively available in other bases such as Web of Science and DBLP are excluded from the sample. Data extraction processed was limited to available articles and pdf conversion readability. Since each publisher has its own text template, data cleaning and text mining processes had increased in complexity. Text mining was performed on English-only article corpus. Such limitations can be surpassed with the addition of other languages' dictionaries, improving semantic broadness.

\paragraph{}Future studies should target on big data privacy cultural aspects. User behavior, laws and regulations, and visual privacy are interesting topics that appeared on this analysis. Studies related to data breaches and practice versus theory evaluation on privacy governance would also be an interesting field to explore. Too little is known about privacy law and regulation causes and effects on people, organizations and government. These studies should be evaluated on their ``intention to inform and evaluate" big data privacy practical effects. Would be desirable that these studies describe the big data privacy implications versus measurable protection practices, their benefits to policymakers, planners, researchers and citizens.

\end{document}